\documentclass[copyright,creativecommons]{eptcs}
\providecommand{\event}{ThEdu 2024}

\usepackage{underscore}
\usepackage[T1]{fontenc}
\usepackage{subcaption}
\usepackage{graphicx}
\usepackage{amsmath}
\usepackage{amssymb}
\usepackage{cite}
\usepackage{cleveref}
\usepackage{tikz}
\usepackage{tabularx}

\RequirePackage{array}
\newenvironment{authors}[1]%
  {\begingroup
   \newcommand\estyle{}%
   \renewcommand\institute[1]%
     {\\\multicolumn{#1}{@{}c@{}}{\scriptsize\begin{tabular}[t]{@{}>{\footnotesize}c@{}}##1\end{tabular}}}%
   \renewcommand\email[1]%
     {\gdef\estyle{\footnotesize\ttfamily}\\##1\gdef\estyle{}}
   \begin{tabular}[t]{@{}*{#1}{>{\estyle}c}@{}}
  }%
  {\end{tabular}%
   \endgroup
  }

\title{Minimal Sequent Calculus for Teaching First-Order Logic: Lessons Learned}

\author{
  \begin{authors}{1}
    Jørgen Villadsen
      \institute{Technical University of Denmark, Kongens Lyngby, Denmark}
      \email{jovi@dtu.dk}
  \end{authors}
}

\def\authorrunning{J. Villadsen}
\def\titlerunning{Minimal Sequent Calculus for Teaching First-Order Logic: Lessons Learned}

\usepackage{latexsym}

\DeclareUnicodeCharacter{00A4}{\raisebox{.08ex}{$\scriptstyle\:\text{\textcurrency}\:$}}
\DeclareUnicodeCharacter{03BB}{\mbox{$\lambda$}}
\DeclareUnicodeCharacter{2192}{\mbox{$\rightarrow$}}
\DeclareUnicodeCharacter{219D}{\mbox{$\leadsto$}}
\DeclareUnicodeCharacter{21D2}{\mbox{$\Rightarrow$}}
\DeclareUnicodeCharacter{2200}{\mbox{$\forall$}}
\DeclareUnicodeCharacter{2203}{\mbox{$\exists$}}
\DeclareUnicodeCharacter{2208}{\mbox{$\in$}}
\DeclareUnicodeCharacter{2227}{\mbox{$\wedge$}}
\DeclareUnicodeCharacter{2261}{\mbox{$\equiv$}}
\DeclareUnicodeCharacter{2286}{\mbox{$\subseteq$}}
\DeclareUnicodeCharacter{22A2}{\mbox{$\vdash$}}
\DeclareUnicodeCharacter{22A5}{\mbox{$\bot$}}
\DeclareUnicodeCharacter{22A8}{\mbox{$\models$}}
\DeclareUnicodeCharacter{22A9}{\mbox{$\Vdash$}}
\DeclareUnicodeCharacter{22C0}{\mbox{$\bigwedge$}}
\DeclareUnicodeCharacter{27F6}{\mbox{$\longrightarrow$}}
\DeclareUnicodeCharacter{27F9}{\mbox{$\Longrightarrow$}}
\DeclareUnicodeCharacter{2AA2}{\raisebox{.25ex}{$\scriptstyle>\!>$}}

\begin{document}
\maketitle

\begin{abstract}
MiniCalc is a web app for teaching first-order logic based on a minimal sequent calculus. As an option the proofs can be verified in the Isabelle proof assistant. We present the lessons learned using the tool in recent years at our university.
\end{abstract}

\section{Introduction}\label{sec:intro}

We present MiniCalc, a web app for teaching first-order logic, based on a so-called minimal sequent calculus.
We explain the sequent calculus in Section~\ref{sec:minimal}.
More than 100 computer science students have used versions of MiniCalc in a course on automated reasoning in the period 2021-2024.

The web app MiniCalc 1.0 has not yet been announced, but it is available here:
\begin{center}
\medskip
\url{https://proof.compute.dtu.dk/MiniCalc.zip}
\medskip
\end{center}
Installation is easy: Just unpack \texttt{MiniCalc.zip} in a new directory and open \texttt{index.html} in a browser.
\begin{center}
\medskip
\includegraphics[width=.95\textwidth]{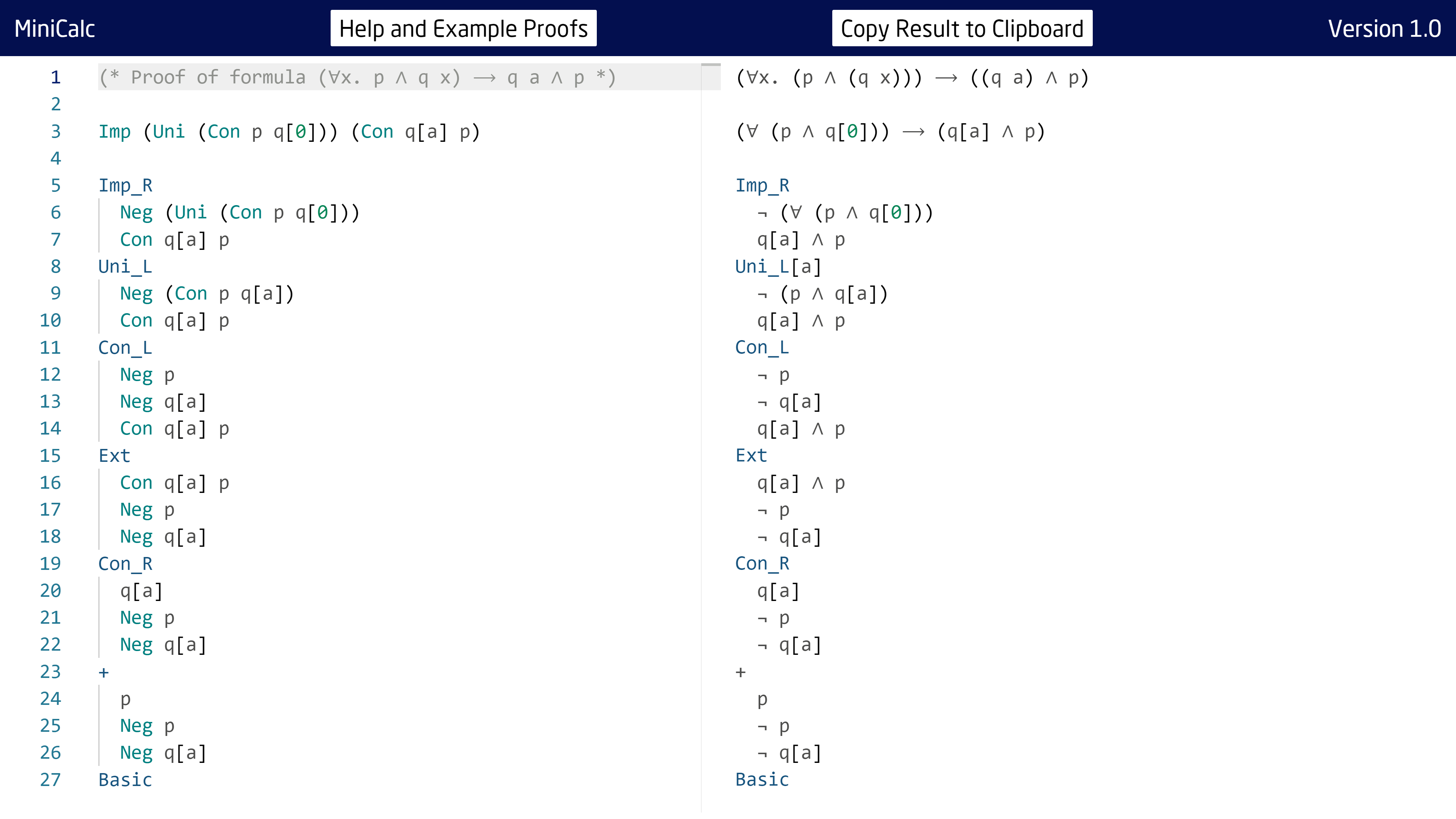}
\medskip
\end{center}
MiniCalc displays the proof editor to the left and the result about the default example proof to the right.
We explain the default example proof in Section~\ref{sec:examples}.
The files in the above zip are from 12 February 2024 and we are not aware of bugs as of 1 December 2024.

We are reluctant to make MiniCalc easily available on the internet, because the final exam in our course is without internet access, and we fear that students will forget to download the zip before entering the exam room, if they can work during the whole course using MiniCalc on the internet.

We also provide a short MiniCalc tutorial (6 pages):
\begin{center}
\medskip
\url{https://proof.compute.dtu.dk/MiniCalc.pdf}
\medskip
\end{center}
Part of it is more of a reference manual, where the full syntax for the proofs in MiniCalc is specified.
We have experienced that computer science students benefit from this information in the early stages of learning to use MiniCalc.
The tutorial also explains how to use the MiniCalc web app to prove formulas in first-order logic and then formally verify the proofs in the Isabelle proof assistant \cite{Isabelle02}.
We follow another style of tutorial in the present paper and in particular we do not assume that MiniCalc has been installed and the tutorial has been studied.

We emphasize the following points:
\begin{enumerate}
\item MiniCalc is not for absolute beginners. Users should have at least an elementary understanding of logic, preferably propositional logic and a bit of first-order logic as well as mathematical maturity. Our experience is that knowledge of functional and/or logic programming is an advantage.
\item MiniCalc can be used without the Isabelle proof assistant, but the formal verification in Isabelle is highly recommended, and the formalization in Isabelle of the syntax, semantics and minimal sequent calculus for first-order logic is a distinguishing feature of MiniCalc.
\item MiniCalc is purposely designed so that there is quite a lot of typing in order to enter the proofs. For instance, the name of the sequent calculus rules must always be stated. Several programming languages are also quite verbose, often enabling better warnings and error messages.
\end{enumerate}

After installation, we suggest the following activities:
\begin{enumerate}
\item
Click on the \textsf{Help and Example Proofs} button for the online help page:
\begin{itemize}
\item How the result can be formally verified in Isabelle
\item How to write a proof in MiniCalc
\begin{itemize}
\item Terms: functions, constants and variables
\item Formulas: predicates, connectives and quantifiers
\item The rules of MiniCalc
\end{itemize}
\item Additional example proofs.
\end{itemize}
The online help page describes almost all of the MiniCalc features, but we have experienced that it can be a bit overwhelming.
\item
Click on the \textsf{Copy Result to Clipboard} button for the result as Isabelle lines.
Follow the above instructions on the online help page if Isabelle is available, or simply paste the Isabelle lines in the MiniCalc editor to have a quick look.
Recall that the formal verification in Isabelle is optional and is not the focus of the present paper at all.
In our automated reasoning course we only accept the resulting Isabelle files for the weekly exercises, the mandatory assignments and the final exam.
\end{enumerate}

In January 2025 we will update the online help file and in particular the tutorial for the 2025 edition of our automated reasoning course.
We constantly make sure that MiniCalc is in sync with the current Isabelle release.

We explain the minimal sequent calculus in Section~\ref{sec:minimal} and example proofs in Section~\ref{sec:examples}.
We consider related work in Section~\ref{sec:related}.
We describe our approach to teaching and lessons learned in Section~\ref{sec:lessons}.

\newpage

\section{Minimal Sequent Calculus}\label{sec:minimal}

We have developed a variant of sequent calculus \cite{Szabo69,DBLP:journals/logcom/FromSV23,DBLP:conf/itp/FromJ22,BenAri2012} that we call minimal due to the concise rules to follow.
We consider first-order logic with functions, that is, not only with constants.
We consider only classical logic here, neither intuitionistic nor minimal logic, although we usually explain these logics elsewhere to students who take our course.

In MiniCalc, we do not want to have to keep track of the names of variables, so instead of identifying variables by conventional names, we will identify them by numbers. The numbering system we use for variables is a bit complicated, since we need to keep track of the quantifier that binds each variable. To do this we use so-called de Bruijn indices to identify the variables.

The idea is to number the variables from the inside out so that the variable with index 0 is always bound by the innermost quantifier, and so on. Using de Bruijn indices, the formula ∀x. ∀y. ∀z. P(x) → Q(y) → R(z) is written ∀ ∀ ∀ (P(2) → Q(1) → R(0)).

Note that the same index may be used for different variables if we have multiple branches within the formula, just like how we may usually use the same variable name in several places as long as their scopes do not overlap. Consider for example the formula ∃x. (∀y. P(y) ∧ Q(x)) → (∀y. Q(x) ∧ P(y)) which is written ∃ (∀ (P(0) ∧ Q(1))) → (∀ (Q(1) ∧ P(0))) using de Bruijn indices.

As mentioned previously, the reason for using de Bruijn indices is that we do not want to keep track of variable names. One reason for this is that the semantics of the MiniCalc system depend on being able to identify variables. An interpretation in MiniCalc contains an environment which maps variables to elements of the domain. The environment changes whenever we encounter a quantifier, since the newly quantified variable must be added to the environment. If we had used usual variable names, we would have to keep track of all of the names that are currently defined to construct this mapping. Since we use de Bruijn indices to represent variables, we can simply increment the index of every existing variable by one to make room for the new variable as the new innermost variable with index 0. Once we leave the scope of a quantifier, the variable it binds no longer exists, we decrement the index of every other variable to make the new innermost variable return to index 0. The environment is then simply a mapping from the natural numbers to elements of the domain.

Another benefit of using de Bruijn indices is that they make it very easy to identify equivalent formulas containing quantifiers. Consider for example the formulas ∀x. P(x) and ∀y. P(y). These formulas clearly have exactly the same meaning, but they are not syntactically identical, so we would need some rule that allows us to prove that they are equivalent. With de Bruijn indices however, both of these formulas are written ∀ P(0), so there is no need to introduce an additional rule to prove the equivalence between the formulas. This allows us to easily prove formulas such as (∀x. P(x)) → (∀y. P(y)).

The universal quantification ∀ A is written Uni A, while the existential quantification ∃ A is written Exi A, and similarly for negation (Neg), implication (Imp), disjunction (Dis) and conjunction (Con). There are 7 kinds of rules:
\begin{enumerate}
\item A basic excluded middle sequent
\item Introduction of implication and negation of implication
\item Introduction of disjunction and negation of disjunction
\item Introduction of conjunction and negation of conjunction
\item Introduction of existential quantifier and negation of existential quantifier
\item Introduction of universal quantifier and negation of universal quantifier
\item Elimination of one extra copy of a formula in a sequent
\end{enumerate}
Our minimal sequent calculus has a very simple structural rule, but the usual structural rules as well as a rule for double negation can be derived in Isabelle.

We elaborate on the rules in 3 steps.

\subsection{Step 1: Naming of Rules}

The sequent calculus is one-sided, using just a single list of formulas as a sequent, which is to be understood as a disjunction of the formulas in the list.

\begin{verbatim}
  Basic: A basic excluded middle sequent
  Imp_R: Introduction of implication
  Imp_L: Introduction of negation of implication
  Dis_R: Introduction of disjunction
  Dis_L: Introduction of negation of disjunction
  Con_R: Introduction of conjunction
  Con_L: Introduction of negation of conjunction
  Exi_R: Introduction of existential quantifier
  Exi_L: Introduction of negation of existential quantifier
  Uni_R: Introduction of universal quantifier
  Uni_L: Introduction of negation of universal quantifier
  Extra: Elimination of one extra copy of a formula in a sequent
\end{verbatim}

We keep the usual left / right distinction where the left rules are rather for the negated formulas.

\subsection{Step 2: Adding Datatypes and Functions}

The syntax is like datatypes in functional programming, no infix notation and using de Bruijn indices for the quantifiers (the students find this quite straight-forward, but they are also advanced computer science students).

\begin{verbatim}
  Basic: ⊩ p # z if member (Neg p) z
  Imp_R: ⊩ Imp p q # z if ⊩ Neg p # q # z
  Imp_L: ⊩ Neg (Imp p q) # z if ⊩ p # z and ⊩ Neg q # z
  Dis_R: ⊩ Dis p q # z if ⊩ p # q # z
  Dis_L: ⊩ Neg (Dis p q) # z if ⊩ Neg p # z and ⊩ Neg q # z
  Con_R: ⊩ Con p q # z if ⊩ p # z and ⊩ q # z
  Con_L: ⊩ Neg (Con p q) # z if ⊩ Neg p # Neg q # z
  Exi_R: ⊩ Exi p # z if ⊩ subt t p # z
  Exi_L: ⊩ Neg (Exi p) # z if ⊩ Neg (inst c p) # z and news c (p # z)
  Uni_R: ⊩ Uni p # z if ⊩ inst c p # z and news c (p # z)
  Uni_L: ⊩ Neg (Uni p) # z if ⊩ Neg (subt t p) # z
  Extra: ⊩ z if ⊩ p # z and member p z
\end{verbatim}

We use \verb|⊩| for the sequent calculus proof system, since \verb|⊢| is used in the Isabelle formalization for a natural deduction proof system, making it possible to discuss both proof systems in the course.

In Isabelle, \verb|#| is used to separate the head and the tail of a list, so we have a very simple criteria for axioms (Basic): the head of the list must have its negation as a member of the tail of the list.

The functions \verb|member| as well as \verb|news| (fresh constant in formulas), \verb|subt| (substitution with term in formula) and \verb|inst| (instantiation with constant in formula) are defined in the Isabelle formalization (again the students find this quite straight-forward).

\subsection{Step 3: Adding Isabelle Notions and Notations}

The sequent calculus is inductively defined in the Isabelle formalization:

\begin{verbatim}
inductive sequent_calculus :: ‹fm list ⇒ bool› (‹⊩ _› 0) where
  Basic: ‹⊩ p # z› if ‹member (Neg p) z› |
  Imp_R: ‹⊩ Imp p q # z› if ‹⊩ Neg p # q # z› |
  Imp_L: ‹⊩ Neg (Imp p q) # z› if ‹⊩ p # z› and ‹⊩ Neg q # z› |
  Dis_R: ‹⊩ Dis p q # z› if ‹⊩ p # q # z› |
  Dis_L: ‹⊩ Neg (Dis p q) # z› if ‹⊩ Neg p # z› and ‹⊩ Neg q # z› |
  Con_R: ‹⊩ Con p q # z› if ‹⊩ p # z› and ‹⊩ q # z› |
  Con_L: ‹⊩ Neg (Con p q) # z› if ‹⊩ Neg p # Neg q # z› |
  Exi_R: ‹⊩ Exi p # z› if ‹⊩ subt t p # z› |
  Exi_L: ‹⊩ Neg (Exi p) # z› if ‹⊩ Neg (inst c p) # z› and ‹news c (p # z)› |
  Uni_R: ‹⊩ Uni p # z› if ‹⊩ inst c p # z› and ‹news c (p # z)› |
  Uni_L: ‹⊩ Neg (Uni p) # z› if ‹⊩ Neg (subt t p) # z› |
  Extra: ‹⊩ z› if ‹⊩ p # z› and ‹member p z›
\end{verbatim}

The quotes \verb|‹...›| are standard in Isabelle formalizations.
We have the following definitions:
\begin{verbatim}
  ‹member p [] = False›
  ‹member p (q # z) = (if p = q then True else member p z)›
\end{verbatim}

\begin{verbatim}
  ‹ext y [] = True›
  ‹ext y (p # z) = (if member p y then ext y z else False)›
\end{verbatim}

The function \verb|ext| (extend) means that all formulas of the second argument must be a member of the first argument, hence we obtain the following derived rule:

\begin{verbatim}
theorem Ext: ‹⊩ y› if ‹⊩ z› and ‹ext y z›
\end{verbatim}

The following 4 small lemmas in Isabelle should help with the intended use of the \verb|Ext| rule:

\begin{verbatim}
lemma ‹ext z z› ‹ext (p # z) z› ‹ext [a,b,c] [c,b]› ‹ext [a,b,c] [a,a,a,b]›
  by auto
\end{verbatim}

Corresponding structural rules:
\begin{itemize}
\item Do nothing, cf.\ \verb|ext z z|
\item Simple weakening (head of list), cf.\ \verb|ext (p # z) z|
\item Permutation (and weakening), cf.\ \verb|ext [a,b,c] [c,b]|
\item Contraction (and weakening), cf.\ \verb|ext [a,b,c] [a,a,a,b]|
\end{itemize}

We can also derive the following rule for double negation:

\begin{verbatim}
theorem NegNeg: ‹⊩ Neg (Neg p) # z› if ‹⊩ p # z›
\end{verbatim}

\section{Example Proofs}\label{sec:examples}

Using the rules of Section~\ref{sec:minimal} we promote the following layout for proofs with line numbers added for ease of reference:

\begin{verbatim}
       1        Imp p p
       2
       3        Imp_R
       4          Neg p
       5          p
       6        Ext
       7          p
       8          Neg p
       9        Basic
\end{verbatim}

However, any layout will do, even everything on a single line, but the promoted layout is always used by MiniCalc, and it supports easy copy-and-paste of a block of lines when entering the proof.
In the Isabelle result the promoted layout is also available as a comment, so it is easy to obtain the promoted layout from any other layout.

The layout has a number of advantages.
Firstly, we see two sequents, in lines 4--5 and in lines 7--8.
Technically there is a also a sequent in line 1, but we always start with a one-formula sequent, and we never need to consider a one-formula sequent again, because we need excluded middle to finish the proof like in line 9.
Note that \verb|Basic| will always be the last line, but \verb|Basic| can also be used elsewhere.
Secondly, all rules work on the head of the list making up the sequent, namely line 4 and line 7, and these lines are easy to spot.

The first line shows that we are about to prove $p → p$. The second line is blank to separate the formula from the proof. We then use the rules \verb|Imp_R|, \verb|Ext| and \verb|Basic|.

The main Isabelle result is as follows with essentially the same layout:
\begin{center}
\includegraphics[width=.35\textwidth]{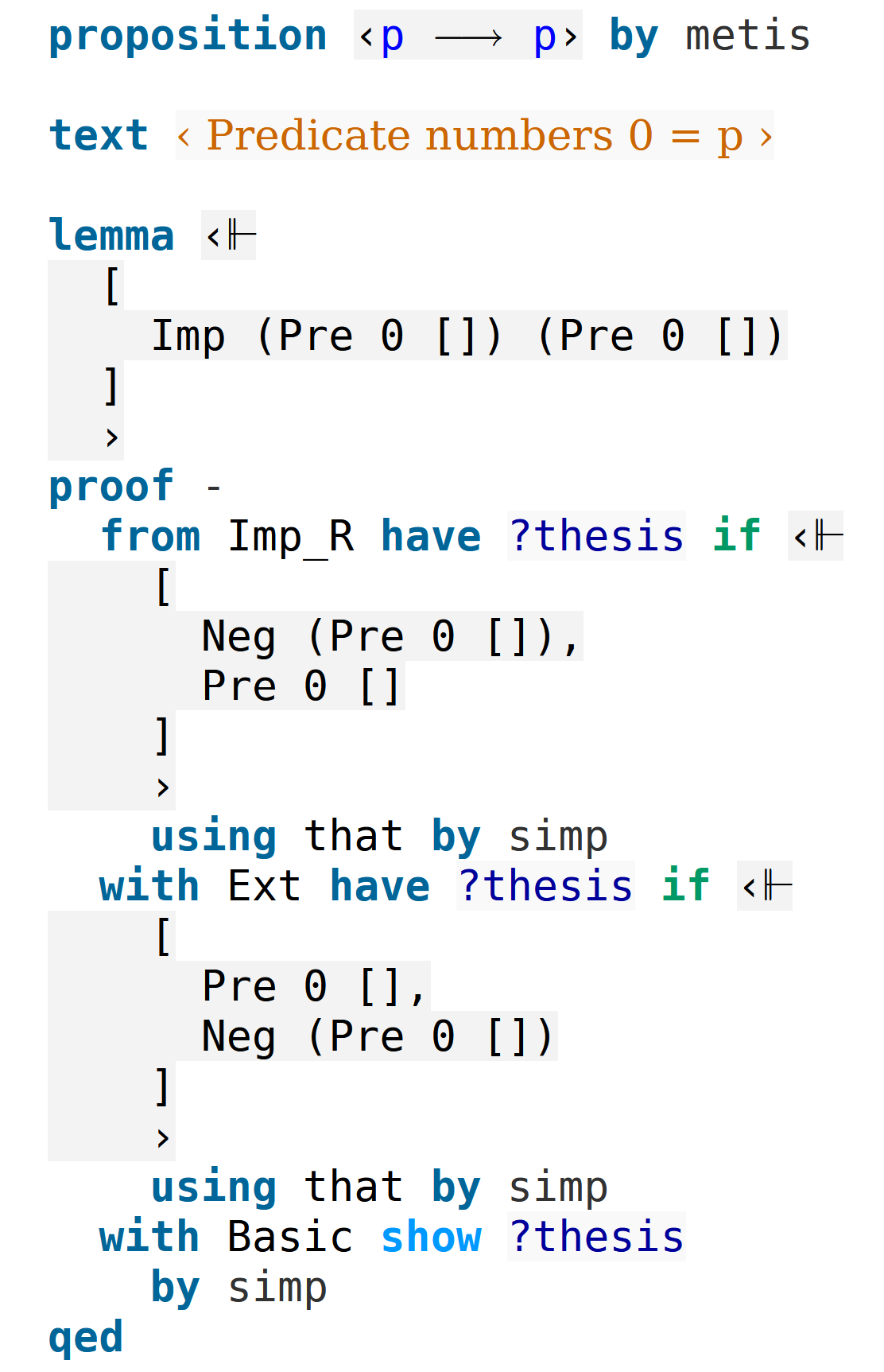}
\end{center}
As mentioned the formal verification in Isabelle is not the focus of the present paper.

\newpage

\subsection{A Trivial Example Proof in First-Order Logic}

The formula $p(a,b) → p(a,b)$ in first-order logic has the same proof as the formula $p → p$ above:

\begin{center}
\includegraphics[width=\textwidth]{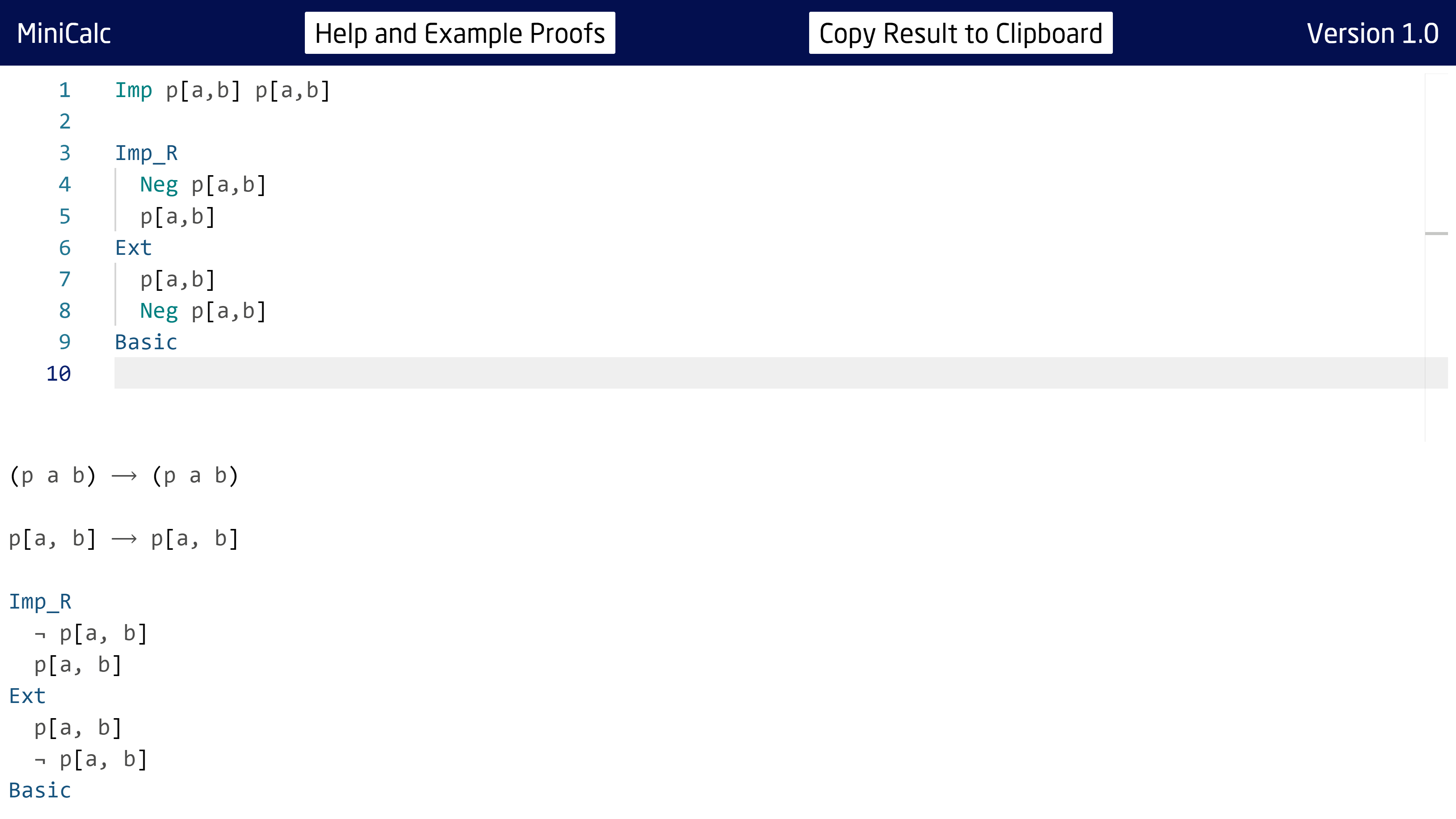}
\end{center}

Note that MiniCalc has the output below or to the right of the editor, depending on the size of the window.

Finally, below is an error message to the user due to missing the proof step with the rule \verb|Ext|:

\begin{center}
\includegraphics[width=\textwidth]{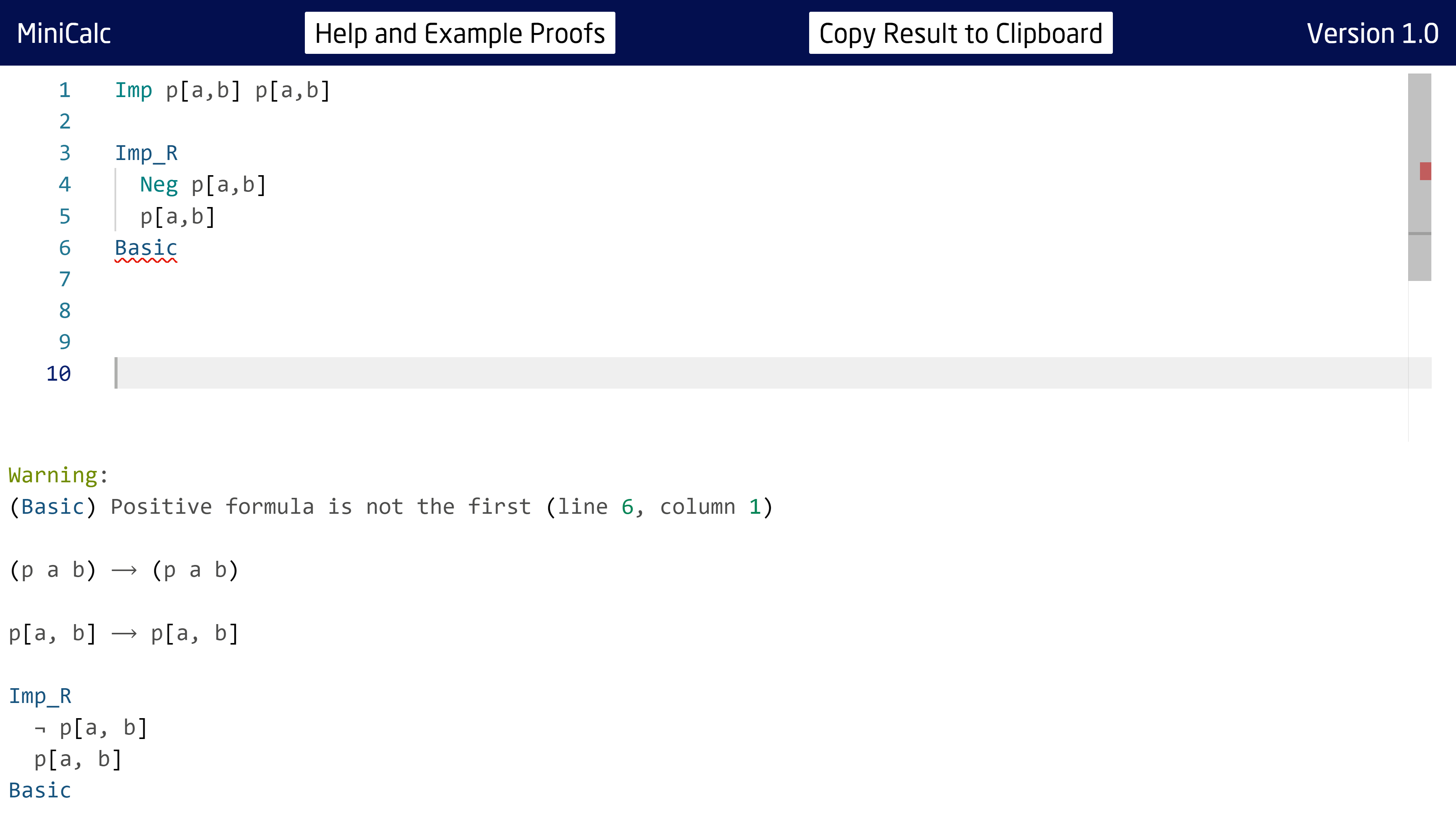}
\end{center}

In the MiniCalc web app it is called a warning, since there could be a bug in the MiniCalc web app, which Isabelle would catch later.

\subsection{The Default Example Proof in First-Order Logic}

We return to the default example proof in Section~\ref{sec:intro} and zoom in on the left and right parts of the window:
\begin{center}
\bigskip
\includegraphics[trim=20mm 22mm 584mm 42mm,clip,width=.48\textwidth]{start}
\hfill
\includegraphics[trim=510mm 22mm 94mm 42mm,clip,width=.48\textwidth]{start}
\bigskip
\end{center}

The formula is shown in 4 formats in the first lines:
\begin{itemize}
\item As a comment by the student to the left in line 1 (a useful way to provide students with formulas to prove) and to the right automatically with all parentheses (useful for students).
\item As entered by the student to the left in line 13 (text-based with de Bruijn indices) and to the right automatically with the usual logical operators (again useful for the students).
\end{itemize}

The format in line 1 is the same as used in Isabelle's higher-order logic (of course we are only using the first-order fragment).

The proof steps start in lines 5, 8, 11, 15, 19 and 27. Note the branching with \verb|+| in line 23; the \verb|Basic| rule can close both. Note also the instantiation \verb|[a]| automatically added to the right in line 8. The student is free to add it to the left as well.

At first it may seem superfluous to have the proof more or less repeated to the right, but students benefit:
\begin{itemize}
\item When the students build up the proof, they often do not care so much about the layout (spaces and line breaks), but they quickly understand the promoted layout and ultimately adapt.
\item When the students enter the text-based format to the right and see the logical operators, they practice the correspondence between the notion (say, implication) and the notation (\verb|→|).
\end{itemize}

\newpage

\subsection{Advanced Example Proofs in First-Order Logic}

We include two variants of the drinker paradox:
\begin{center}
\medskip
\url{https://en.wikipedia.org/wiki/Drinker_paradox}
\medskip
\end{center}

Note that the initial formula must be duplicated in order to instantiate for $a$ as well as $b$, cf.\ lines 6--7, 8 ($a$) and 17 ($b$):

\begin{center}
\includegraphics[trim=20mm 30mm 734mm 30mm,clip,width=.45\textwidth]{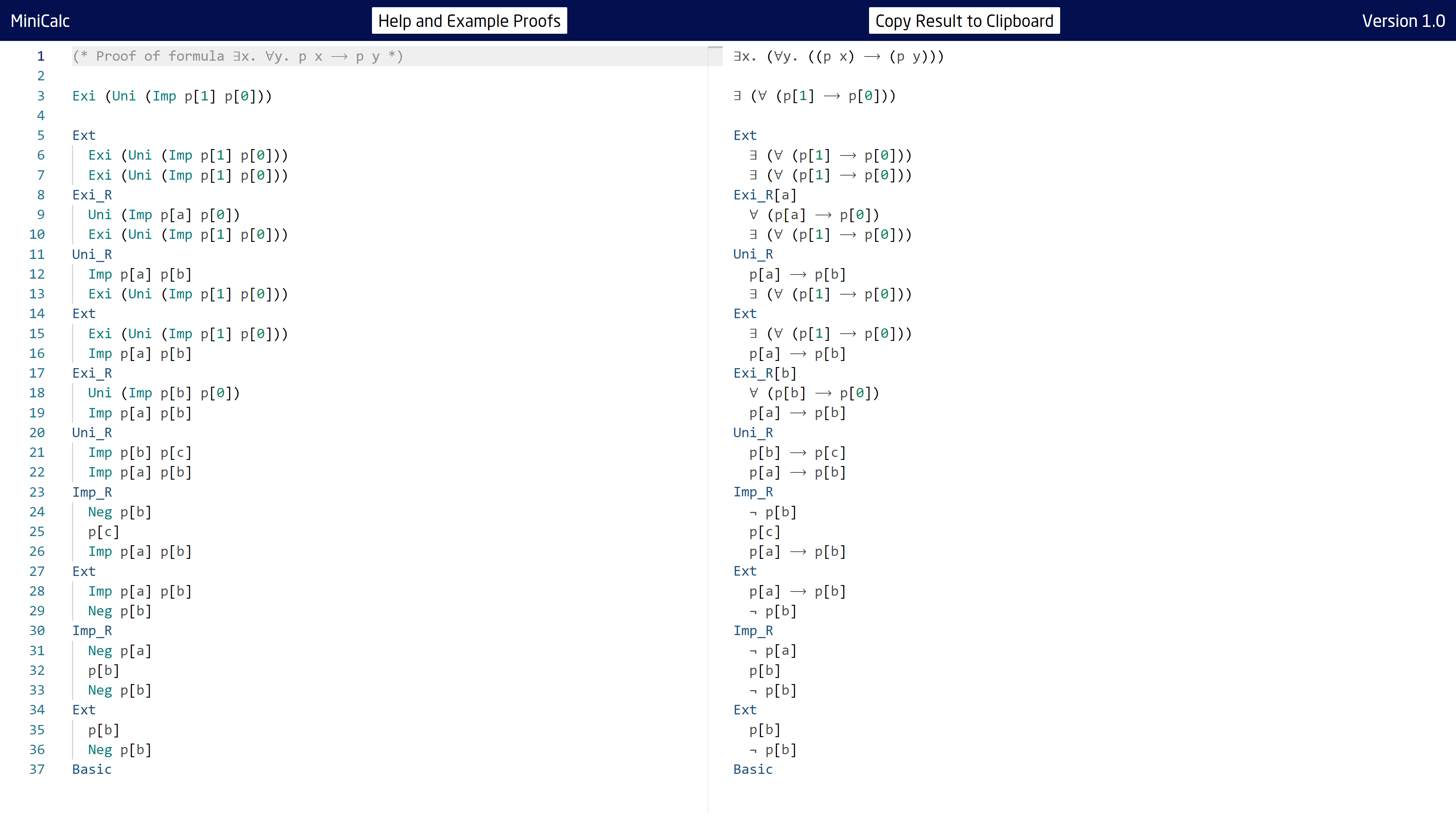}
\hfill
\includegraphics[trim=510mm 30mm 244mm 30mm,clip,width=.45\textwidth]{e1}
\end{center}

\newpage

A shorter proof of the same formula:
\begin{center}
\includegraphics[trim=20mm 30mm 734mm 30mm,clip,width=.45\textwidth]{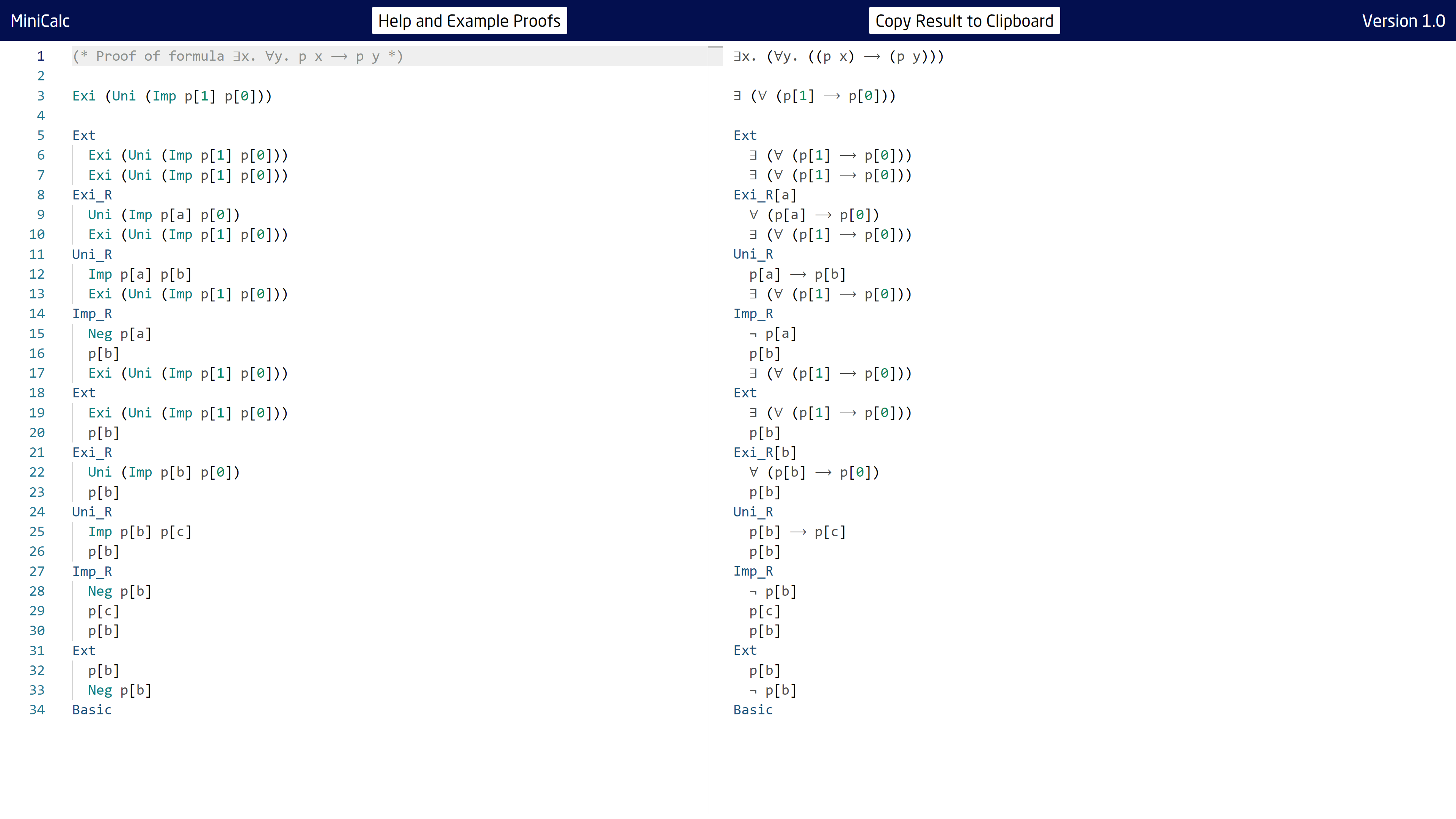}
\hfill
\includegraphics[trim=510mm 30mm 244mm 30mm,clip,width=.45\textwidth]{e2}
\end{center}

\section{Related Work}\label{sec:related}

There are many tools today for teaching first-order logic, but we are only aware of our NaDeA system \cite{nadea} and our SeCaV system \cite{LSFA} where the proof results are checked in Isabelle and furthermore with formalized proofs of soundness and completeness.
Both of these systems concern classical first-order logic with functions.
We compare MiniCalc and SeCaV in the following section.
With respect to NaDeA, it is totally different, natural deduction instead of sequent calculus, and the user interface is by clicking only instead of text-based proofs.
At the moment we do not use NaDeA in our courses, but it is occasionally used elsewhere.

There are other formalizations of sequent calculus in proof assistants \cite{blanchettepopescu,Synthetic-Completeness-AFP,braselmanncomplete,ilik,ilikleeherbelin,persson,herbelinkimlee,10.1007/978-3-031-43513-3-25,DBLP:conf/cade/VilladsenSF18}.
None of these works consider any tools for teaching logic.

\newpage

There are several tools for sequent calculus, like Carnap.io \cite{Carnap17}, Sequoia~\cite{ReisNH20}, Sequent Calculus Trainer~\cite{SCT17} and the Logitext web app:
\begin{center}
\medskip
\url{http://logitext.mit.edu}
\medskip
\end{center}
AXolotl~\cite{Axolotl19,CernaSSWB20} is an Android app that supports sequent calculus and it is designed to facilitate self-study.
Unlike SeCaV, none of these tools provide any formal guarantees of their correctness.

The Incredible Proof Machine~\cite{Breitner16} takes proofs to be so-called port graphs, with a model of this proof idea formalized in Isabelle and shown to be as strong as natural deduction.
Unlike SeCaV, the formalized metatheoretical results only apply to a model of the system instead of to the actual implementation.

Our Natural Deduction Assistant (NaDeA)~\cite{nadea} presents natural deduction in a more traditional style.
Its metatheory is formalized in Isabelle and the web app supports exporting proofs that can be verified in Isabelle, alleviating the problem of potential bugs:
\begin{center}
\medskip
\url{https://nadea.compute.dtu.dk/}
\medskip
\end{center}

Our so-called SeCaV Unshortener was very briefly explained some years ago at 16th Logical and Semantic Frameworks with Applications, LSFA 2021, 23-24 July 2021 \cite{LSFA}.
Only a few paragraphs and screenshots for the proof of $p \to p$ (back then \verb'Imp_R' was called \verb'AlphaImp' as common in tableaux proofs).

Our paper \cite{LSFA} described the logical design considerations of SeCaV (Sequent Calculus Verified) as a theory in Isabelle, but the web app is still available:
\begin{center}
\medskip
\url{https://secav.compute.dtu.dk/}
\medskip
\end{center}
In comparison with the SeCaV Unshortener, MiniCalc also has a much improved editor, with colors, folding and underlining of the error, and the present paper is the first presentation of the web app (besides the few paragraphs in the LSFA paper \cite{LSFA}).

\section{Our Approach to Teaching and Lessons Learned}\label{sec:lessons}

We have used the prover as teaching material in our MSc course on automated reasoning \cite{FMTea,ThEdu20}.
The course was given for the first time in 2020.
Our university makes a summary of the course evaluations available online and the course has good evaluations.
The course is an optional 5~ECTS course.
It is important to know that although the course is not mandatory it becomes mandatory for the student once the student registers for the exam.

\begin{center}
\begin{tabular}{l@{\hspace{2em}}r@{\hspace{1em}}r@{\hspace{1em}}r@{\hspace{1em}}r@{\hspace{1em}}r@{\hspace{2em}}r}
Year & 2020 & 2021 & 2022 & 2023 & 2024 & Total \\[1ex]
Number of students registered & 26 & 35 & 40 & 19 & 34 & 154 \\[1ex]
Number of students passed & 24 & 30 & 36 & 15 & 28 & 133
\end{tabular}
\end{center}

\begin{center}
    \begin{tikzpicture}
        \node [rectangle,font=\footnotesize] (Headline) at (0,0) {
        \begin{tabularx}{0.98\textwidth}{>{\centering\arraybackslash\hsize=1.1111\hsize}X>{\centering\arraybackslash\hsize=1.1111\hsize}X>{\centering\arraybackslash\hsize=1.1111\hsize}X>{\centering\arraybackslash\hsize=1.1111\hsize}X>{\centering\arraybackslash\hsize=0.5555\hsize}X}
                \multicolumn{5}{c}{\emph{Selected courses related to automated reasoning in the BSc and MSc computer science programs}} \\[0.01\textwidth]
                \\[0.0001\textwidth] Year 1 & Year 2 & Year 3 & Year 4 & \dots \\[0.008\textwidth]\hline
                \multicolumn{3}{|c|}{BSc} & \multicolumn{2}{c}{\qquad\qquad \ MSc}\\\hline
            \end{tabularx}
        };
        
        \node [rectangle,draw,align=center,minimum width=0.206\textwidth,minimum height=0.09\textwidth,font=\footnotesize] (IP) at (-6.20,-1.9) {Discrete\\Mathematics\\(mandatory)};
        \node [rectangle,draw,align=center,minimum width=0.206\textwidth,minimum height=0.09\textwidth,font=\footnotesize] (DM) at (-6.20,-3.5) {Introductory\\Programming\\(mandatory)};
        \node [rectangle,draw,align=center,minimum width=0.206\textwidth,minimum height=0.09\textwidth,font=\footnotesize] (A1) at (-6.20,-5.1) {Algorithms and\\Data Structures 1\\(mandatory)};
        \node [rectangle,draw,align=center,minimum width=0.206\textwidth,minimum height=0.09\textwidth,font=\footnotesize] (S1) at (-6.20,-6.7) {Software\\Engineering 1\\(mandatory)};
        \node [rectangle,draw,align=center,minimum width=0.206\textwidth,minimum height=0.09\textwidth,font=\footnotesize] (A2) at (-2.75,-1.9) {Computer Science\\Modelling\\(mandatory)};
        \node [rectangle,draw,align=center,minimum width=0.206\textwidth,minimum height=0.09\textwidth,font=\footnotesize] (CS) at (-2.75,-3.5) {Functional\\Programming\\(mandatory)};
        \node [rectangle,draw,align=center,minimum width=0.206\textwidth,minimum height=0.09\textwidth,font=\footnotesize] (FP) at (-2.75,-5.1) {Algorithms and\\Data Structures 2\\};
        \node [rectangle,draw,align=center,minimum width=0.206\textwidth,minimum height=0.09\textwidth,font=\footnotesize] (S2) at (-2.75,-6.7) {Software\\Engineering 2\\};
        \node [rectangle,draw,align=center,minimum width=0.206\textwidth,minimum height=0.09\textwidth,font=\footnotesize] (LSLP) at (0.70,-1.9) {Logical Systems\\and Logic\\Programming};
        \node [rectangle,draw,align=center,minimum width=0.206\textwidth,minimum height=0.09\textwidth,font=\footnotesize] (CC) at (0.70,-3.5) {Concurrent\\Programming\\};
        \node [rectangle,draw,align=center,minimum width=0.206\textwidth,minimum height=0.09\textwidth,font=\footnotesize] (OS) at (0.70,-5.1) {Operating\\Systems\\};
        \node [rectangle,draw,align=center,minimum width=0.206\textwidth,minimum height=0.09\textwidth,font=\footnotesize] (AI) at (0.70,-6.7) {Introduction to\\Artificial\\Intelligence};
        
        \node [rectangle,draw,align=center,minimum width=0.206\textwidth,minimum height=0.09\textwidth,font=\bfseries] (AR) at (4.20,-1.9) {Automated\\Reasoning};
        
        \node [rectangle,draw,align=center,minimum width=0.206\textwidth,minimum height=0.09\textwidth,font=\footnotesize] (LS) at (5.9,-3.5) {Logic for\\Security\\};
        \node [rectangle,draw,align=center,minimum width=0.206\textwidth,minimum height=0.09\textwidth,font=\footnotesize] (AFP) at (5.9,-5.1) {Program\\Verification\\};
        \node [rectangle,draw,align=center,minimum width=0.206\textwidth,minimum height=0.09\textwidth,font=\footnotesize] (FASP) at (5.9,-6.7) {Model\\Checking\\};
        \node [rectangle,draw,align=center,minimum width=0.206\textwidth,minimum height=0.09\textwidth,font=\footnotesize] (LTUL) at (5.9,-8.3) {Formal Aspects\\of Software\\Engineering};
        
        \draw [->] (AR.230) |- (LS.west);
        \draw [->] (AR.230) |- (AFP.west);
        \draw [->] (AR.230) |- (FASP.west);
        \draw [->] (AR.230) |- (LTUL.west);
    \end{tikzpicture}
\end{center}

MiniCalc has been successfully tested in a 5 ECTS optional course Automated Reasoning (one of two courses described elsewhere \cite{FMTea}):
\begin{center}
\medskip
\url{https://courses.compute.dtu.dk/02256/}
\medskip
\end{center}
MiniCalc counts for 30\%\ of the grade.
The remaining 70\%\ of the exam consists of Isabelle/Isar proofs \cite{isar-ref}.

There is a button in MiniCalc that copies the resulting proof to the clipboard and when pasted in Isabelle the proof is formally checked (the student must import a special Isabelle file as described in the tutorial).
However, we have not experienced that the MiniCalc web app has produced a wrong proof, but it is part of the teaching philosophy to verify everything in Isabelle.
Anyway, the present paper describes the web app as a stand-alone tool.

The course Automated Reasoning is an introduction to automatic and interactive theorem proving, and as mentioned Isabelle is used to formalize almost all of the concepts we introduce during the course.

Students end proving the following complicated formula in the course:

\begin{center}
\includegraphics[width=.8\textwidth]{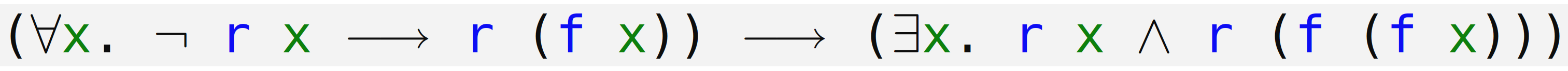}
\end{center}

A possible formulation in English:
If every person that is not rich has a rich father, then some rich person must have a rich grandfather.

We ask students to hand in six assignments during the course, but we also have a two hour written exam at the end of the course, with all aids allowed but no internet. 
We find that this setup works well for our course.

\newpage

The exam consists of three problems, each of which contain 3-4 questions.
Each of the three problems concerns a different topic, namely:
\begin{enumerate}
    \item Sequent calculus proofs --- Using MiniCalc (result checked in Isabelle)
    \item Natural deduction proofs --- Using Isabelle/Pure (no automation)
    \item General proofs in Isabelle/HOL
\end{enumerate}

One of the main benefits of using Isabelle for our exam is of course that much of the grading of the exams is almost automatic.

A key point is that we have a formalization of the syntax, semantics and proof systems.
We also have formal soundness and completeness theorems for the proof systems.
The formalizations are part of the teaching materials. 

\

\section*{Acknowledgements}
Thanks to Frederik Krogsdal Jacobsen and Simon Tobias Lund for help.

\

\bibliographystyle{eptcs}
\bibliography{references}

\end{document}